\documentclass[apj]{emulateapj}

\def\gtorder{\mathrel{\raise.3ex\hbox{$>$}\mkern-14mu
             \lower0.6ex\hbox{$\sim$}}}
\def\ltorder{\mathrel{\raise.3ex\hbox{$<$}\mkern-14mu
             \lower0.6ex\hbox{$\sim$}}}

\slugcomment{Draft of \today}

\shorttitle{SN\,2009ip: progenitor mass-loss rate}

\shortauthors{Ofek et al.}

\begin{document}

\title{SN\,2009ip: Constraints on the progenitor mass-loss rate}
\author{E.~O.~Ofek\altaffilmark{1},
L.~Lin\altaffilmark{2},
C.~Kouveliotou\altaffilmark{3},
G.~Younes\altaffilmark{4},
E.~G\"{o}\u{g}\"{u}\c{s}\altaffilmark{2},
M.~M.~Kasliwal\altaffilmark{5},
Y.~Cao\altaffilmark{6}}

\altaffiltext{1}{Benoziyo Center for Astrophysics and the Helen Kimmel center for planetary science, Weizmann Institute of Science, 76100 Rehovot, Israel.}
\altaffiltext{2}{Faculty of Engineering and Natural Sciences, Sabanc\i~University, Orhanl\i $-$ Tuzla, \.{I}stanbul 34956, Turkey}
\altaffiltext{3}{Space Science Office, ZP-12, NASA/Marshall Space Flight Center, Huntsville, AL 35812, USA}
\altaffiltext{4}{Universities Space Research Association, 6767 Old Madison Pike NW, Suite 450, Huntsville, AL 35806, USA}
\altaffiltext{5}{Observatories of the Carnegie Institution for Science, 813 Santa Barbara St, Pasadena CA 91101.}
\altaffiltext{6}{Division of Physics, Mathematics and Astronomy,
  California Institute of Technology, Pasadena, CA 91125.}

\begin{abstract}


Some supernovae (SNe) show evidence for
mass-loss events taking place prior to their explosions.
Measuring their pre-outburst mass-loss rates
provide essential information regarding the mechanisms
that are responsible for these events.
Here we present {\it XMM}-Newton and {\it Swift} X-ray
observations taken after the latest,
and presumably the final, outburst of SN\,2009ip.
We use these observations as well as
new near infra-red and visible light spectra,
and published radio and visible light observations
to put six independent
order-of-magnitude constrains on the mass-loss rate of the SN progenitor
prior to the explosion.
Our methods utilize:
the X-ray luminosity, the bound-free absorption,
the H$\alpha$ luminosity, the SN rise-time,
free-free absorption, and the bolometric luminosity of the outburst
detected prior to the explosion.
Assuming spherical mass-loss with a wind density profile,
we estimate that the effective mass-loss rate
from the progenitor was between
$10^{-3}$ to $10^{-2}$\,M$_{\odot}$\,yr$^{-1}$,
over a few years prior to the explosion,
with a velocity of $\sim10^{3}$\,km\,s$^{-1}$.
This mass-loss rate corresponds to a total 
circum stellar matter mass of $\sim0.04$\,M$_{\odot}$,
within $6\times10^{15}$\,cm of the SN.
We note that the mass-loss rate estimate based on
the H$\alpha$ luminosity is higher by an order of magnitude.
This can be explained if the narrow line H$\alpha$
component is generated at radii larger than the shock radius,
or if the CSM has an aspherical geometry.
We discuss simple geometries which are
consistent with our results.

\end{abstract}


\section{Introduction}
\label{sec:Introduction}

Supernova (SN) observations,
especially of Type IIn (e.g., Filippenko 1997),
indicate that some
massive stars lose considerable amounts of mass
($\gtorder 10^{-4}$\,M$_{\odot}$) within a few months to
years prior to their explosions
(e.g., Dopita et al. 1984;
Chugai et al. 1994, 2004;
Ofek et al. 2007, 2010, 2013b;
Smith et al. 2007, 2008, 2009;
Kiewe et al. 2012).
Several theoretical mechanisms to eject large amounts of
mass
with super-Eddington
luminosities have been suggested.
Quataert \& Shiode (2012) suggest that in some massive
stars the super-Eddington fusion luminosities,
shortly prior to core collapse,
can drive convective motions, that in turn excite
gravity waves that propagate toward the stellar surface.
The dissipation of these waves can unbind up to several
solar masses of the stellar envelope.
In Ofek et al. (2013b) we argued that this mechanism
can unbind a lower amount of mass
($\sim 10^{-2}$\,M$_{\odot}$).
Arnett \& Meakin (2011) suggested that shell oxygen burning in
massive stars produces large fluctuations in the turbulent
kinetic energy, that in turn may produce bursts.
Chevalier (2012) suggested that the mass loss
is driven by a common-envelope phase
due to the inspiral of a neutron star into a giant companion core,
unbinding the companion envelope and setting up accretion
onto the neutron star, that in turn collapses into a black hole and
triggers a SN explosion.
Soker \& Kashi (2013) suggested that the SN\,2009ip explosion
was due to the merger of two stars, while some of the pre-explosion
outbursts occurred near periastron passages of the binary system.
Another possible mechanism
is the pulsational pair instability
which in very massive stars can generate
several explosions, expelling $\gtorder1$\,M$_{\odot}$ each,
followed by the collapse of the stellar core
(Rakavy, Shaviv \& Zinamon 1967;
Woosley, Blinnikov \& Heger 2007;
Waldman 2008).

Measuring the mass-loss rates from massive stars
prior to their explosion
can be used as a tool to study the latest stages of stellar evolution,
and to discriminate between the different models suggested to
generate large mass-loss events.
Objects in which super-Eddington outbursts were directly observed prior to
the SN explosion provide a way to constrain the time
at which mass-loss was taking place, 
and relate the optical luminosities
with mass-loss rates and kinetic energy estimates.
To date there are only three SNe in which precursor
outbursts were seen prior to the SN explosion.
These are the Type Ibn
SN\,2006jc (e.g., Foley et al. 2007; Pastorello et al. 2008),
the Type IIn SN\,2009ip (e.g., Mauerhan et al. 2012;
Pastorello et al. 2012;
Prieto et al. 2012)
and the Type IIn SN\,2010mc/PTF\,10tel (Ofek et al. 2013b).

Here we present {\it XMM}-Newton and {\it Swift}
X-ray observations of SN\,2009ip.
We use these observations as well as 
published and new visible light and radio observations,
to set an order of magnitude estimate 
on the mass-loss prior to the SN explosion.

SN\,2009ip was a Luminous Blue Variable (LBV)
originally detected in outburst
on 2009 August 26.11 by the CHASE survey
at a projected distance of 4.3\,kpc
from NGC\,7259 (Maza et al. 2009).
Three additional outbursts were subsequently discovered with the
Catalina Real-Time Transient Survey
on 2010 July 15, on 2010 September 29 (Drake et al. 2010),
and then again on 2012 July 24 (Drake et al. 2012).
Based on its multiple outbursts,
Smith et al. (2010) and Foley et al. (2011)
argued that it is a supernova impostor
(see recent reviews in Kochanek, Szczygie{\l}, \& Stanek 2012;
Smith et al. 2011; van Dyk \& Matheson 2012).
In September 2012, Smith \& Mauerhan (2012), and later Mauerhan et al. (2012),
reported the detection of broad P-Cygni lines with velocities of up to
13,000\,km\,s$^{-1}$,
suggesting that the star had finally exploded as a Type IIn SN.
Previous cases in which a likely LBV progenitor has exploded
as a supernova include SN\,2005gl
(Gal-Yam et al. 2007; Gal-Yam \& Leonard 2009),
and SN\,1961V (Kochanek, Szczygiel, \& Stanek 2011; Smith et al. 2011).
Prieto et al. (2012) reported that around 2012 September 24
the object's $I$-band light curve started to rise rapidly
at a rate of 2.3\,mag\,day$^{-1}$.
Shortly afterwards, on early October, X-ray emission
was detected from SN\,2009ip with the {\it Swift}/X-Ray Telescope
(XRT; Margutti \& Soderberg 2012b).

Throughout the paper we assume that the source is located
at a distance of 20.4\,Mpc.
In \S\ref{sec:Obs} we present our 
observations of SN\,2009ip,
while in \S\ref{sec:Form} we review
various methods for estimating the mass
content of the SN circumstellar matter (CSM).
Finally, in \S\ref{sec:Disc} we apply these
methods to SN\,2009ip and discuss our findings.

\section{Observations}
\label{sec:Obs}


We observed SN\,2009ip with {\it XMM}-Newton on 2012 November 1
in prime full window imaging mode for an effective exposure time of 8\,ks. Using data collected
with the EPIC--pn detector (Tian et al. 2007), we accumulated the source
spectrum from a circular region of 30$\arcsec$ centered on the optical
position of SN\,2009ip. We selected a circular background region from a
source free area on the same chip (i.e., CCD 7) with the same
aperture size. The source is detected at a significance of about
3\,$\sigma$ with a background subtracted count rate of
(4.6$\pm$1.5)$\times$10$^{-3}$ counts\,s$^{-1}$, yielding a total of
37 net source counts in the 0.5$-$10 keV. We generated the detector
and ancillary response files using the latest calibration data.

\begin{figure}
\centerline{\includegraphics[width=8.5cm]{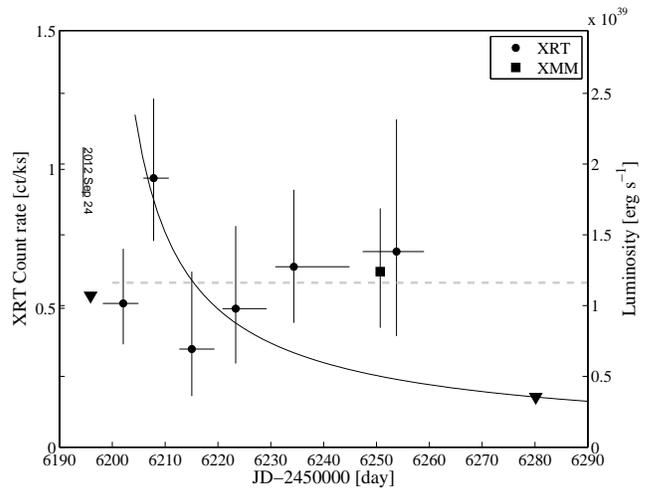}}
\caption{X-ray light curve of SN\,2009ip based on {\it Swift}--XRT
(circles) and {\it XMM}-Newton (square) observations.
The triangles mark XRT 2-$\sigma$ upper limit.
The horizontal error bars represent the range of observations
in each bin.
The gray dashed line indicates the mean XRT count rates level
of the observations taken between 2012 September 29
and 2012 Nov 28.
It corresponds to a luminosity of
$\approx1.1\times10^{39}$\,erg\,s$^{-1}$.
We note that the left-hand axis count rate corresponds
only to the {\it Swift}--XRT observations.
The right-hand axis shows the unabsorbed luminosity
assuming,
a Galactic hydrogen column density of
$N_{{\rm H}}=1.2\times10^{20}$\,cm$^{-2}$,
and an X-ray spectrum of the form
$n(E)\propto E^{-1.8}$, where $n(E)$ is the photon
numbers per unit energy.
The black solid line shows the expected, order of magnitude,
evolution of the
X-ray luminosity assuming optically thin wind-profile CSM with
mass-loss rate of $7\times10^{-4}$\,M$_{\odot}$\,yr$^{-1}$ and
$v_{w}=500$\,km\,s$^{-1}$ (based on Eqs.~\ref{Lx} and \ref{r}).}
\label{fig:SN2009ip_XLC}
\end{figure}

The {\it Swift}--XRT (Gehrels et al. 2004)
observed SN\,2009ip on an almost daily basis since
2012 September 4 (triggered by Roming/Maragutti).
Some of these X-ray observations have been already reported
in e.g., Margutti et al. (2012), and Campana (2012).
For each {\em Swift}--XRT image of the SN,
we extracted the number of X-ray counts in the 0.2--10\,keV
band within an aperture of $9''$ radius
centered on the SN position.
We note that this aperture contains $\approx 50$\% of
the source flux (Moretti et al.\ 2004).
The background count rates were estimated in
annuls around the SN location, with an inner (outer) radius of $50''$ ($100''$).
The log of {\it Swift}-XRT observations,
along with the source and background X-ray counts in the individual
observations are listed in Table~\ref{tab:XRTobs}.
SN\,2009ip is only marginally detected in individual images,
but it is clearly visible in the coadded data.
Figure~\ref{fig:SN2009ip_XLC} shows a binned light curve
based on the {\it Swift}--XRT
observations.
The binned measurements are listed
in Table~\ref{tab:XRTbin}.
\begin{figure*}
\centerline{\includegraphics[width=6.1cm]{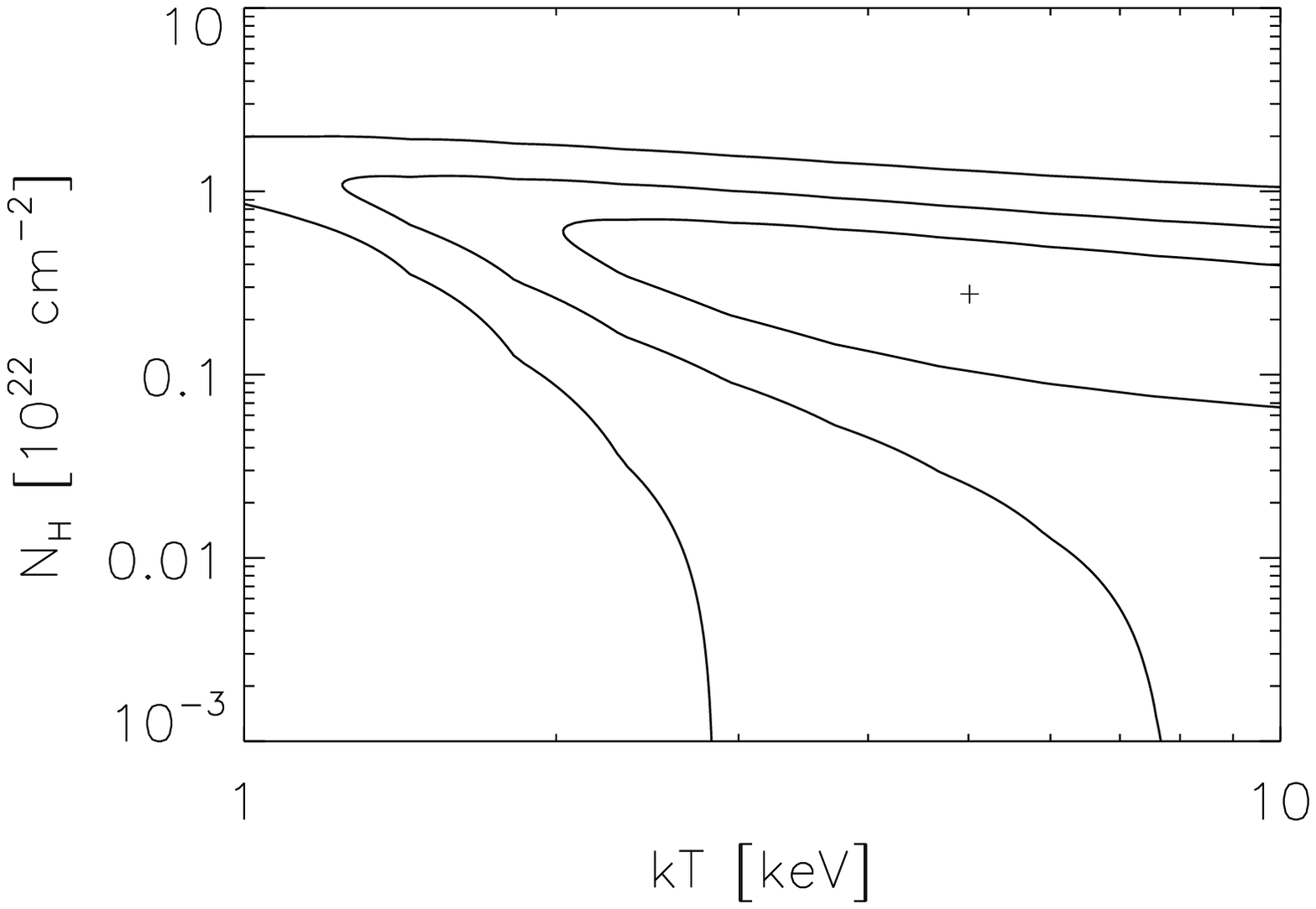}
\includegraphics[width=6.1cm]{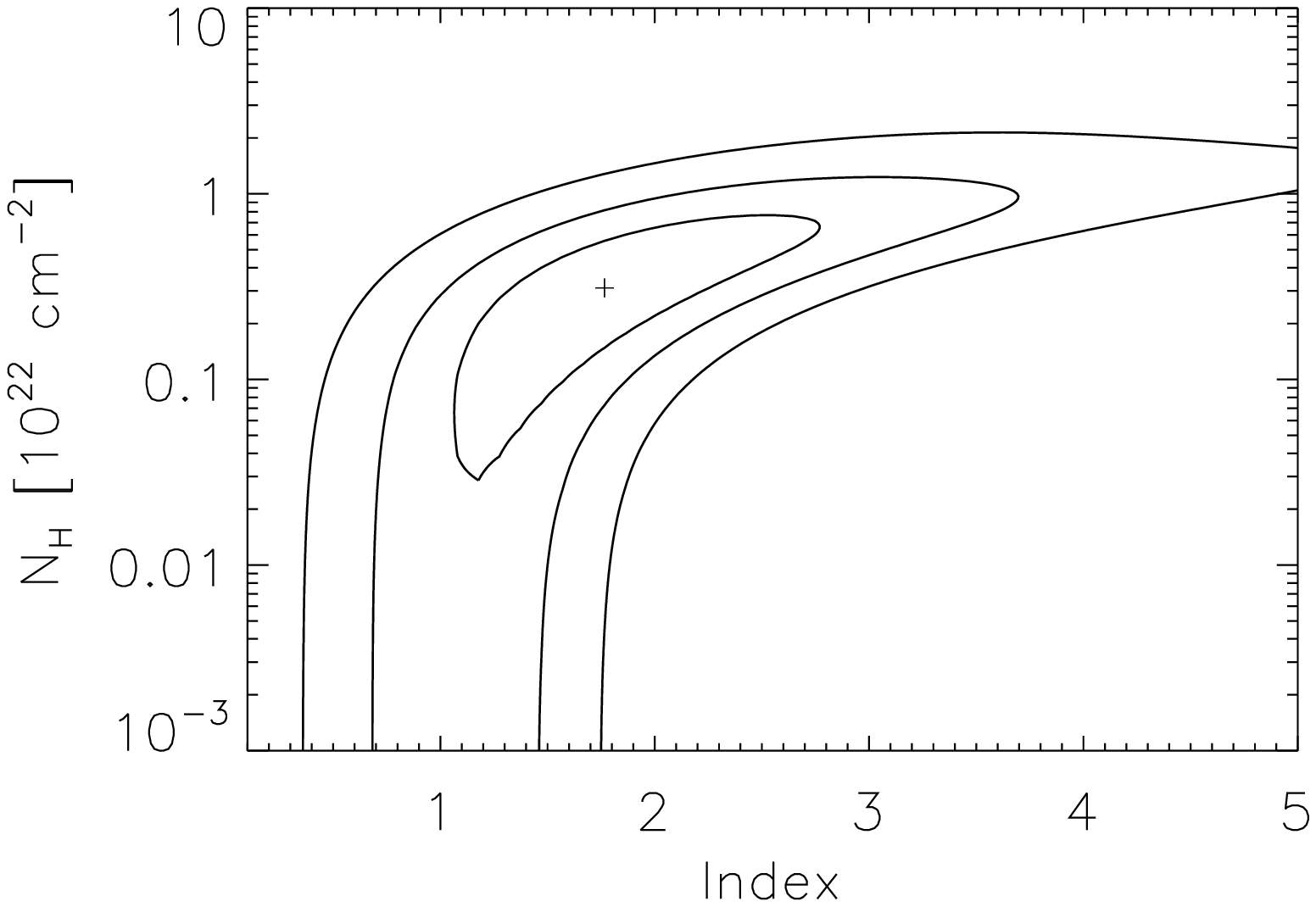}
\includegraphics[width=6.1cm]{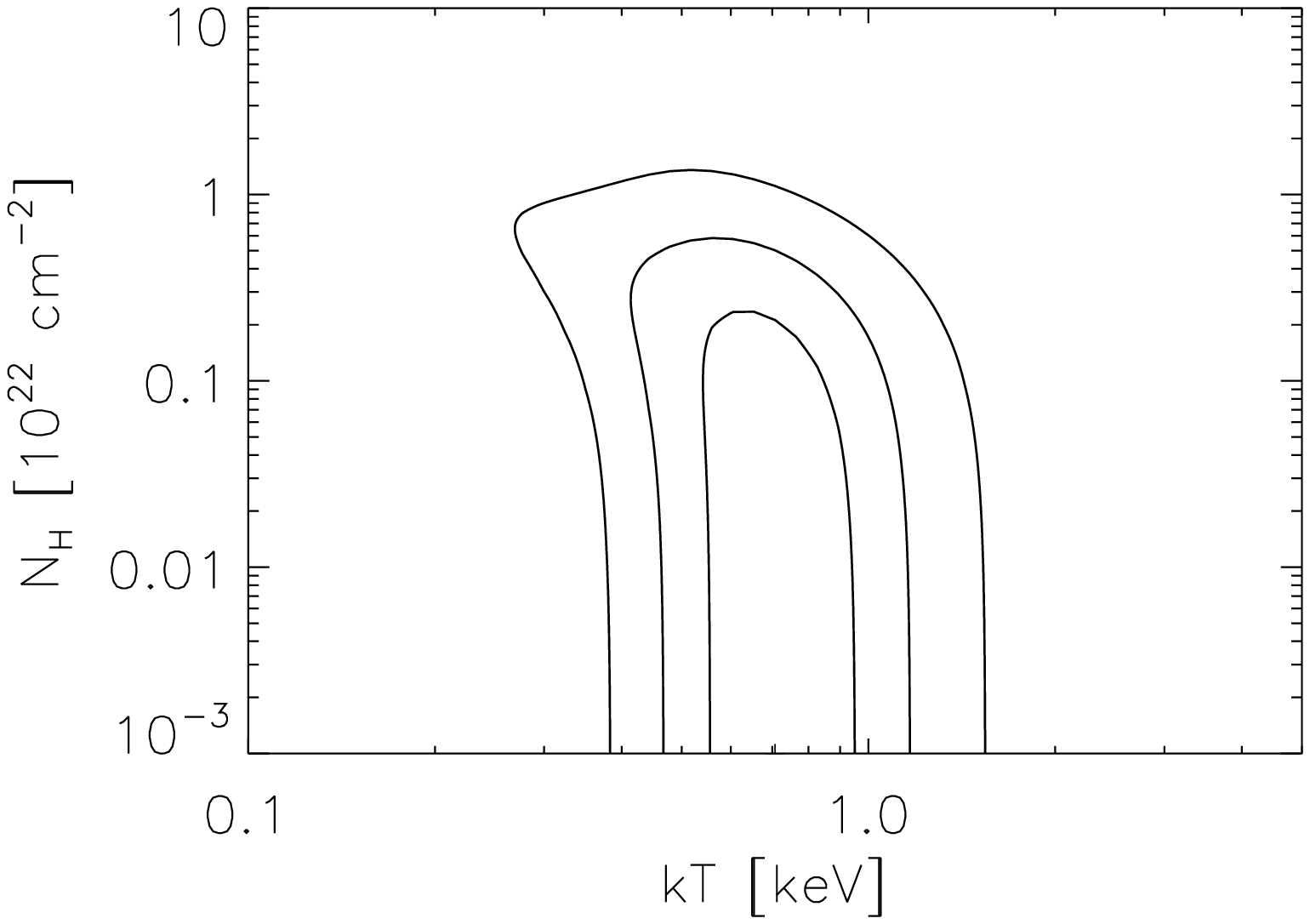}}
\caption{{\bf Left:} $\Delta\chi^{2}$ contours for fitting
the {\it XMM} and {\it Swift}--XRT spectra with a Mekal
model in the column density vs. temperature ($kT$) space
(see text for details).
The contours represent the 1, 2 and 3-$\sigma$ errors,
while the plus sign represents the best fit model.
{\bf middle:} The same as the left panel, but for a power-law model.
The X-axis represents the power law index, $\Gamma$.
{\bf Right:} The same as the left panel, but for a black-body model.
The X-axis represents the black-body temperature in keV.}
\label{fig:Xspec}
\end{figure*}
\begin{deluxetable}{llll}
\tablecolumns{4}
\tablewidth{0pt}
\tablecaption{{\it Swift}-XRT observations}
\tablehead{
\colhead{MJD}          &
\colhead{Exposure time}          &
\colhead{Source}      &
\colhead{Background}     \\
\colhead{day} &
\colhead{ks} &
\colhead{ct}            &
\colhead{ct}
}
\startdata
55084.44  & 9.86 &  0 & 19 \\
56174.86  & 1.96 &  0 &  4 \\
56176.60  & 1.78 &  0 &  4 \\
56183.40  & 1.65 &  0 &  4 \\
56190.75  & 0.39 &  0 &  0 
\enddata
\tablecomments{MJD is the modified Julian day.
Source is the number of counts in 
a $9''$-radius aperture of the source position
and in the 0.2--10\,keV band.
Background, is the number of counts,
in the 0.2--10\,keV band,
in an annuls of inner (outer)
radius of $50''$ ($100''$) around the source.
The ratio between the background annulus area and the aperture area
is $92.59$.
This table is published in its entirety in the electronic edition of
{\it ApJ}. A portion of the full table is shown here for
guidance regarding its form and content.}
\label{tab:XRTobs}
\end{deluxetable}
\begin{deluxetable}{llllll}
\tablecolumns{6}
\tablewidth{0pt}
\tablecaption{{\it Swift}-XRT binned data}
\tablehead{
\colhead{$\langle{\rm MJD}\rangle$} &
\multicolumn{2}{c}{Range}      &
\colhead{CR}                 &
\colhead{UL$_{2\sigma}$}        &
\colhead{Exp.}                \\
\colhead{day}                 &
\colhead{day}                 &
\colhead{day}                 &
\colhead{cnt/ks}       &
\colhead{cnt/ks}       &
\colhead{ks}                
}
\startdata
 55084.4& $-0.0$&$ 0.0$ &  \nodata           &  0.61 &   9.86\\
 56178.0& $-3.2$&$ 12.$7&  \nodata           &  1.03 &   5.79\\
 56195.4& $-2.7$&$ 2.0$ &  \nodata           &  0.55 &  17.40\\
 56201.6& $-3.9$&$ 2.9$ &  $0.52_{-0.15}^{+0.20}$&\nodata&  46.38\\
 56207.3& $-1.9$&$ 2.9$ &  $0.97_{-0.23}^{+0.29}$&\nodata&  37.17\\
 56214.5& $-2.3$&$ 4.3$ &  $0.35_{-0.17}^{+0.28}$&\nodata&  22.66\\
 56222.8& $-2.5$&$ 5.9$ &  $0.50_{-0.20}^{+0.30}$&\nodata&  24.08\\
 56233.9& $-3.5$&$ 10.5$&  $0.65_{-0.20}^{+0.28}$&\nodata&  30.80\\
 56253.3& $-6.4$&$ 5.2$ &  $0.70_{-0.30}^{+0.48}$&\nodata&  14.20\\
 56279.7& $-19.4$&$14.7$&  \nodata            & 0.18  &  52.57
\enddata
\tablecomments{Binned {\it Swift}-XRT light curve of SN\,2009ip.
$\langle{\rm MJD}\rangle$ is the weighted mean modified Julian day
of all the observations in a given bin,
where the observations are weighted by their exposure times.
Range is the time range around $\langle{\rm MJD}\rangle$
in which the light curve
(Table~\ref{tab:XRTobs}) was binned.
CR is the counts rate along
with the lower and upper 1-$\sigma$ errors.
We note that the source count rates are corrected
for extraction aperture losses.
UL$_{2\sigma}$ is the 2-$\sigma$ upper limit on the source count
rate, which is given if the total source counts within the binned
exposure is $\le1$. Exp is the exposure time.}
\label{tab:XRTbin}
\end{deluxetable}

For our XRT spectral analysis we selected
all the XRT observations 
between MJD 56174 and 56228,
taken in photon counting mode and with an integration longer than 500\,s.
This resulted in a total effective exposure time
of 149\,ks.
We extracted a stacked source spectrum from a circular region centered at
the SN location with a radius of 20$\arcsec$. The stacked background
spectrum was accumulated from a 20$\arcsec$ circular source-free region
for all observations. We grouped the source spectrum with a minimum of
10 counts in each energy bin. The background subtracted count rate of
the source is $(4.1\pm0.7)\times10^{-4}$\,counts\,s$^{-1}$,
corresponding to 61 net source counts.

We used {\tt XSPEC}\footnote{http://heasarc.gsfc.nasa.gov/xanadu/xspec/}
V12.7.1 (Schafer 1991) to simultaneously fit the
{\it XMM} and {\it Swift}--XRT spectra.
In all cases we set the Galactic extinction
to $N_{{\rm H}}=1.2\times10^{20}$\,cm$^{-2}$
(Dickey \& Lockman 1990),
and we fitted four parameters:
normalization of the {\it XMM} spectrum,
normalization of the {\it Swift}--XRT spectrum,
a parameter describing the spectrum
(i.e., temperature or power-law index),
and the hydrogen column density at the redshift
of the supernova ($z=0.00594$) assuming solar metallicity.
The best fit parameters
are listed in Table~\ref{tab:Xspec}.
\begin{deluxetable}{llll}
\tablecolumns{4}
\tablewidth{0pt}
\tablecaption{X-ray spectral parameters}
\tablehead{
\colhead{Model}          &
\colhead{parameter}      &
\colhead{$N_{{\rm H}}$}  &
\colhead{$\chi^{2}$/d.o.f.} \\
\colhead{}               &
\colhead{}               &
\colhead{cm$^{-2}$}      &
\colhead{}               
}
\startdata  
Mekal         & $kT=4.74_{-2.3}^{+18}$\,keV   & $(2.8_{-1.6}^{+2.4})\times10^{21}$  & $14.66/13$ \\
Power law     & $\Gamma=1.79_{-0.50}^{+0.60}$  & $(3.2_{-2.0}^{+2.7})\times10^{21}$ & $14.98/13$ \\
Black body    & $kT=0.72\pm0.10$\,keV       & \nodata                         & $18.19/13$ 
\enddata
\tablecomments{$\Gamma$ is defined as the power-law index
in a spectrum of the form, $n(E)\propto E^{-\Gamma}$,
where $n(E)$ is the number of photons
per unit energy. d.o.f. is the number of degrees of freedom.}
\label{tab:Xspec}
\end{deluxetable}
Figure~\ref{fig:Xspec} shows the $\Delta\chi^{2}$
contours of these fits in the
$N_{{\rm H}}$ vs. temperature ($kT$) or power-law index space.
From these fits we can set a $3$-$\sigma$ upper limit on
$N_{{\rm H}}$ in the SN CSM of $2\times10^{22}$\,cm$^{-2}$.


On 2012 September 22,
two days before the fast rise in the light curve,
we obtained a near-infrared (NIR) spectrum of SN\,2009ip
with the Folded-port InfraRed Echellette spectrograph (FIRE;
Simcoe et al. 2008, 2010) on the 6.5-m Magellan Baade Telescope. We
used the low-dispersion, high-throughput prism mode, and completed an
ABBA dither sequence.
The data span 0.8--2.5\,$\mu$m at a resolution ranging from 300--500.
Immediately afterwards, we obtained a spectrum of an A0V standard
star for the purposes of flux calibration and removal of telluric
absorption features, as described by Vacca, Cushing, \& Rayner (2003).
Data were reduced using the FIREHOSE pipeline developed by R. Simcoe,
J. Bochanski, and M. Matejek.
Smith et al. (2013) present detailed analysis of the NIR spectrum.

\begin{figure}
\centerline{\includegraphics[width=8.5cm]{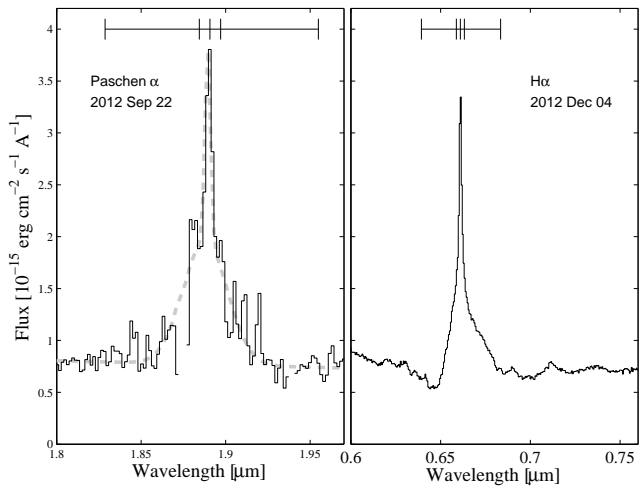}}
\caption{{\bf Left:} Near IR spectrum of SN\,2009ip.
The X-axis is the wavelength at the observer's frame.
The spectrum is centered on the Paschen\,$\alpha$
line. The vertical lines on the scale corresponds 
(from left to right) to velocities of
$-10^{4}$, $-10^{3}$, $0$, $10^{3}$ and $10^{4}$\,km\,s$^{-1}$
relative to the line center.
The dashed-gray line shows the best-fit two-Gaussian model
(we note that the Gaussian were convolved
with the approximate instrumental broadening).
The gaps in the plot are due to the removal of bad/noisy pixels.
{\bf Right:} Same as the left spectrum, but for the visible-light
H$\alpha$ line.}
\label{fig:Spec_Ha_PaschenAlpha}
\end{figure}

On 2012 December 4, we obtained a visible light spectrum
of SN\,2009ip using the
Dual Imaging Spectrograph (DIS) mounted on the ARC-3.5m telescope
at Apache Point Observatory to get a spectrum,
with 600\,s integration, of SN2009ip in the wavelength 
range from 3500\AA~to 9000\AA~and resolution of about 400.
The visible-light spectrum was flux calibrated
using the standard star BD$+28^{\circ}4211$
Parts of the IR and visible light spectra,
centered on the Paschen\,$\alpha$ and H$\alpha$ lines, respectively,
are shown in Figure~\ref{fig:Spec_Ha_PaschenAlpha}.
The full spectra are available from
the WISeREP archive
(Yaron \& Gal-Yam 2012).
We fitted a two Gaussians model to the
Paschen\,$\alpha$ and H$\alpha$ lines.
We find that in the Paschen\,$\alpha$,
the narrow (wide) component width corresponds
to velocity of $\approx200$ ($\approx2100$)\,km\,s$^{-1}$.
In the H$\alpha$ line,
the narrow line component width corresponds
to velocity of $\approx300$\,km\,s$^{-1}$,
while the difference between the emission line center
and the bottom of the P-Cygni absorption feature is
about 8000\,km\,s$^{-1}$.
The flux of the H$\alpha$ narrow component is about
$3\times10^{-14}$\,erg\,cm$^{-2}$\,s$^{-1}$.
Since we do~not have access to photometric measurements
of the SN obtained around the same time in which we got
the visible light spectrum, we estimate that the line flux measurement
is good to about 30\%. For future calibration,
we note that, based on the current calibration,
the SDSS AB synthetic magnitudes of the visible
light spectrum are 17.17, 16.31 and 16.11 in the $g$, $r$ and $i$-bands,
respectively.

\section{Mass loss estimators}
\label{sec:Form}

Here we review several methods that we use to constrain
the mass-loss rate from SN progenitors.
In \S\ref{sec:Disc} we implement these methods for the case of SN\,2009ip.

We use several observables, including
the bound-free absorption limit derived from the X-ray spectrum,
the X-ray luminosity,
upper limit on the diffusion time scale as derived from
the SN rise time,
the H$\alpha$ luminosity,
the non detection in radio bands,
and the bolometric luminosity of the precursor observed
prior to the SN explosion.

Our model assumes that the interaction of the SN blast wave
with the CSM produces X-ray and radio emission at the radius
of the shock. The visible light photons may be produced
below, or at, the shock radius.
The H$\alpha$ emission can be emitted either at the shock
region, if it is due to collisional processes,
or above the shock radius if it originates from optically
thin CSM ionized by the SN radiation field.
All the photon diffusion and attenuation is taking place
above the shock.

Throughout the paper 
we assume that the
CSM around the progenitor
has a spherical wind-density profile of the form
$\rho=Kr^{-2}$, where
$r$ is the distance from the progenitor,
$K\equiv \dot{M}/(4\pi v_{{\rm w}})$ is the mass-loading parameter,
$\dot{M}$ is the mass-loss rate,
and $v_{{\rm w}}$ is the wind/outburst velocity.
Given the outbursts observed in SN\,2009ip
prior to its final explosion,
it is likely that the CSM was~not ejected
as a continuous wind with a uniform velocity.
However, the mass-loss rate estimators
we use below, are not very sensitive to this
assumption.
The reason for this is that,
for a reasonable density distribution,
the emission, or attenuation,
are calculated by integrals
which are dominated by the value
at the shock radius.
Here, the only exception is the mass-loss estimator
based on the H$\alpha$ line luminosity (see \S\ref{sec:Ha}).
Therefore, we argue that the use of the continuous
wind density profile provides an order of magnitude
estimate for the mass-loss rate.
In the following sections we discuss our specific
mass-loss rate estimators and their caveats.

\subsection{Bound free absorption}
\label{sec:bf}

The particle density profile, in continuous wind,
is given by (e.g., Chevalier 1982)
\begin{eqnarray}
n      & \approx   & \frac{1}{\langle \mu_{{\rm p}}\rangle} \frac{\dot{M}}{4\pi m_{{\rm p}} v_{{\rm w}}r^{2}} \cr
       & \cong     & 6\times10^{8} \frac{1}{\langle \mu_{{\rm p}}\rangle} \dot{M}_{0.01} v_{{\rm w},500}^{-1} r\
_{15}^{-2}\,{\rm cm}^{-3},
\label{eq:n}
\end{eqnarray}
where $\dot{M}_{0.01}$ is the mass-loss rate in units of $10^{-2}$\,M$_{\odot}$\,yr$^{-1}$,
$v_{{\rm w},500}$ is the wind/outburst speed in units of 500\,km\,s$^{-1}$,
$r_{15}$ is the radius in units of $10^{15}$\,cm,
$m_{{\rm p}}$ is the proton mass and
$\langle \mu_{{\rm p}}\rangle$ is the mean number of nucleons per particle
(mean molecular weight).
For our order-of-magnitude calculation, we adopt
$\langle \mu_{{\rm p}}\rangle=0.6$.
In a wind profile, the column density between
radius $r$ and infinity is
\begin{equation}
N   = \int_{r}^{\infty}{ndr} \approx 1\times10^{24} \dot{M}_{0.01} v_{{\rm w},500}^{-1} r_{15}^{-1}\,{\rm cm}^{-2}.
\label{N}
\end{equation}

Assuming that the gas in the pre-shocked wind is neutral
and has solar abundance,
the bound-free optical depth in the 0.03--10\,keV region
is roughly given by
(e.g., Behar et al.\ 2011)\footnote{This approximation deviates by a factor of two from a more accurate calculation (e.g., Morrison \& McCammon 1983).}
\begin{eqnarray}
\tau_{{\rm bf}} & =       &  N\sigma(E) \cr
             & \approx & 60 \dot{M}_{0.01} v_{{\rm w},500}^{-1} r_{15}^{-1} E_{1}^{-2.5},
\label{tauX}
\end{eqnarray}
where $\sigma(E)$ is the bound-free cross section as a function
of energy $E$, and $E_{1}$ is the energy in keV.
This approximation is valid when the material
is neutral.
However, since above $\sim0.5$\,keV,
metals with a high ionization potential
dominate the absorption,
this formula is still valid, to an order of a magnitude,
above 0.5\,keV when some of the inner electrons
of the metals are bound (i.e., even if all the hydrogen is ionized).
Chevalier \& Irwin (2012) estimated that
the metals will be completely ionized only
above shock velocities of about $10^{4}$\,km\,s$^{-1}$.

\subsection{H$\alpha$ luminosity}
\label{sec:Ha}

Assuming the SN radiation field can ionize
all the hydrogen in the CSM,
the mass of the hydrogen generating the
H$\alpha$ line is
\begin{equation}
M_{{\rm H}} \approx \frac{m_{{\rm p}} L_{{\rm H}\alpha}}{h\nu_{{\rm H}}\alpha_{{\rm H}}^{{\rm eff}} n_{{\rm e}}}.
\label{eq:MH}
\end{equation}
Here $h$ is the Planck constant,
$L_{{\rm H}\alpha}$ is the Balmer H$\alpha$
line luminosity,
$\nu_{{\rm H}}$ is the line frequency
($4.57\times10^{14}$\,Hz for H$\alpha$),
and $\alpha_{{\rm H}}^{{\rm eff}}$ is the case-B effective
recombination coefficient at 10,000\,K
($\approx8.7\times10^{-14}$\,cm$^{3}$\,s$^{-1}$ for H$\alpha$;
Osterbrock \& Ferland 2006).
An important caveat is that this estimate assumes
that the line is generated by recombination.
Any H$\alpha$ radiation generated in the shocked region
due to collisional processes, is not included here.
In order to avoid this problem we take as the line luminosity
only the luminosity of the narrow line component,
which we assume is due to wind above the shock region.
In a wind profile, the integrated mass from radius $r$ to $r_{1}$
is
\begin{eqnarray}
M & = & \int_{r}^{r_{1}}{4\pi r^{2} Kr^{-2}dr}=4\pi K(r_{1}-r) \sim 4\pi K\beta r \cr
  & \cong & 0.006 \beta \dot{M}_{0.01} v_{{\rm w},500}^{-1} r_{15}\,{\rm M}_{\odot},
\label{eq:Mr}
\end{eqnarray}
where $\beta\equiv (r_{1}-r)/r$.
We note that $\beta$ cannot be arbitrarily large
(otherwise the mass in the CSM will diverge),
and here we will assume it is of order unity.

By substituting Equation~\ref{eq:n}
into Equation~\ref{eq:MH} (assuming $n\approx n_{{\rm e}}$)
and setting it equal to Equation~\ref{eq:Mr},
we can get a relation between the mass-loading parameter $K$,
the H$\alpha$ luminosity, and the radius (Ofek et al. 2013b)
\begin{eqnarray}
L_{{\rm H}\alpha} & \ltorder & \frac{4\pi h \nu_{{\rm H}} \alpha^{{\rm eff}} \beta K^{2}}{ \langle \mu_{{\rm p}}\rangle m_{{\rm p}}^{2} r } \cr
               & \approx  & 2\times10^{39} \dot{M}_{0.01}^{2} v_{{\rm w},500}^{-2} \beta r_{15}^{-1}\,{\rm erg\,s}^{-1},
\label{eq:KLr}
\end{eqnarray}
The reason for the inequality is that it is possible that not
all of the hydrogen is ionized or that the temperature of the gas
is higher than $10^{4}$\,K
(i.e., $\alpha^{{\rm eff}}_{{\rm H}\alpha}$ depends on temperature),
and that $\beta>1$.
We note that if $\beta<1$ then this inequality is incorrect.
However, it is reasonable to assume that the width (i.e., $r_{1}-r$)
of the hydrogen envelope is of the same order of magnitude of $r$.

Another important caveat
(that can be expressed in terms of $\beta$)
is that unlike the
X-ray and radio emission which originate
at the shock region (radius $r$),
the narrow component of the H$\alpha$
may originate at radii $r_{{\rm H}\alpha}>r$ (i.e., above the shock region).
In this case
$L_{{\rm H}\alpha} \propto \dot{M}_{{\rm H}\alpha}^{2}r_{{\rm H}\alpha}^{-1}$,
then $\dot{M}_{{\rm H}\alpha}/\dot{M}\approx (r_{{\rm H}\alpha}/r)^{1/2}$.
Therefore, if $r_{{\rm H}\alpha}$ is an order of magnitude, or more,
larger than $r$, the contribution of $\dot{M}_{{\rm H}\alpha}$
to the bound-free column density
(Eq.~\ref{N}) and the diffusion time scale discussed in \S\ref{sec:tdiff},
will be small.

\subsection{X-ray emission}
\label{sec:x}

The X-ray emission from optically thin
region is given by (e.g., Immler et al. 2008)
\begin{equation}
L_{{\rm X}} \approx \int_{r}^{\infty}{4\pi r^{2} \Lambda(T) n^{2}dr},
\label{eq:Lxn}
\end{equation}
where
$\Lambda(T)$ is the effective cooling function
in the 0.2--10\,keV range.
Assuming an optically thin thermal plasma with a temperature
in the range $10^{6}$--$10^{8}$\,K (Raymond et al. 1976),
we adopt a value of $\Lambda(T)\approx 3\times10^{-23}$\,erg\,cm$^{3}$\,s$^{-1}$.
Substituting Equation~\ref{eq:n} into Equation~\ref{eq:Lxn}
we get (e.g., Ofek et al. 2013b)
\begin{eqnarray}
L_{{\rm X}} & \approx & 4\pi \Lambda(T) \frac{K^{2}}{\langle \mu_{{\rm p}}\rangle^{2} m_{{\rm p}}^{2}r} {\rm e}^{-(\tau+\tau_{{\rm bf}})}\cr
          & \approx   & 3.8\times10^{41} \dot{M}_{0.01}^{2} v_{{\rm w},500}^{-2} r_{15}^{-1} {\rm e}^{-(\tau+\tau_{{\rm bf}})}\,{\rm erg\,s}^{-1}.
\label{Lx}
\end{eqnarray}
This expression includes additional exponential term
due to absorption in the wind,
where $\tau$ is the Thomson optical depth
(see Ofek et al. 2010; Balberg \& Loeb 2011),
which is $\sim 0.3\dot{M}_{0.01} v_{{\rm w},500}^{-1} r_{15}^{-1}$.
Although the Thomson optical depth is well known,
when the optical depth is of the order of a few,
Compton scattering is expected to reprocess more energetic
photons into the 0.2--10\,keV band
(Chevalier \& Irwin 2012; Svirski et al. 2012).
Since the exact X-ray spectrum is not known
(Katz et al. 2011; Svirski et al. 2012), a proper calculation of
$L_{{\rm X}}$ when $\tau\gtorder1$ is not straightforward.
Assuming that
the Comptonization of hard X-rays into the soft X-rays band
is smaller than the reduction of soft X-ray flux
by the optical-depth factors,
Equation~\ref{Lx} provides an order of magnitude lower limit on $\dot{M}$.

In Figure~\ref{fig:SN2009ip_XLC},
the black-solid line shows an order of magnitude estimate of the
expected X-ray luminosity,
assuming an optically-thin wind-profile CSM with
a mass-loss rate of $7\times10^{-4}$\,M$_{\odot}$\,yr$^{-1}$ and
$v_{w}=500$\,km\,s$^{-1}$ (eqs.~\ref{Lx} and \ref{r}).
Some points are discrepant by factors of 2--3 in luminosity
from this estimate.
However, this is a simplistic model
and since $\dot{M} \propto L_{{\rm X}}^{1/2}$, our mass-loss estimate
based on the X-ray luminosity is plausibly correct to within
an order of magnitude.

We conclude that this formula can be trusted
only for $\dot{M}\ltorder 10^{-2}$\,M$_{\odot}$\,yr$^{-1}$.
Above this mass-loss rate, $\tau$ and $\tau_{{\rm bf}}$
are larger than unity.

\subsection{Diffusion time scale}
\label{sec:tdiff}

Another observable that can be used to constrain the mass-loss
rate is the rise time of the SN light curve.
If a considerable amount of material is present
between the SN and the observer,
then photon diffusion will slow down the rise time of the SN light curve.
Therefore, the maximum observed SN rise time can be used to put an upper limit
on the amount of mass between the SN and the observer.
The diffusion time scale in an infinite
wind profile is given by e.g., Ginzburg \& Balberg (2012)
\begin{eqnarray}
t_{{\rm diff}} & \approx &\frac{\kappa K}{c}[\ln{\Big(\frac{c}{v_{{\rm sh}}}\Big)}-1] \cr
             & \cong   & 0.13 \kappa_{0.34} \dot{M}_{0.01} v_{{\rm w},500}^{-1} [\ln(30 v_{{\rm sh},4}^{-1})-1]\,{\rm day}.
\label{tdiff}
\end{eqnarray}
Here $v_{{\rm sh},4}$ is the SN shock velocity in units
of $10^{4}$\,km\,s$^{-1}$.
In the case of SN\,2009ip
the early fast rise of the SN light curve provides
an upper limit on $t_{{\rm diff}}$ and, therefore,
an upper limit on $\dot{M}$.

\subsection{Free-free absorption}
\label{sec:ff}

Typically,
SN progenitors with mass-loss rates of
$\sim10^{-6}$\,M$_{\odot}$\,yr$^{-1}$ are easily detectable in radio frequencies
in the nearby universe
(e.g., Horesh et al.\ 2012;
Krauss et al.\ 2012).
The radio emission is the result of an interaction
between the SN shock and the CSM that generates synchrotron radiation
peaking at radio frequencies
(e.g., Slysh 1990;
Chevalier \& Fransson 1994;
Chevalier 1998).
However, if the material is ionized or partially ionized
then the free-free optical depth may block
this radiation.
The free-free optical depth in a wind profile,
between radius $r$ and the observer,
is given by
(e.g., Ofek et al. 2013a)
\begin{equation}
\tau_{{\rm ff}} \approx 1.0\times10^{5} T_{{\rm e}, 4}^{-1.35} \nu_{10}^{-2.1} v_{{\rm w},500}^{-2} \dot{M}_{0.01}^{2} r_{15}^{-3},
\label{tau_ff}
\end{equation}
where $\nu_{10}$ is the frequency in units of 10\,GHz.
We note that the presence of Balmer lines in the spectrum
likely means that at least some of the Hydrogen is ionized
and therefore, free-free absorption is important.

Chandra \& Soderberg (2012)
and Hancock et al. (2012)
reported on radio observations of SN\,2009ip obtained on 2012 September 26 using the Jansky Very Large Array (JVLA\footnote{The Jansky Very
Large Array is operated by the National Radio Astronomy Observatory (NRAO),
a facility of the National Science Foundation operated under
cooperative agreement by Associated Universities, Inc.}),
and the Australia Telescope Compact Array (ATCA).
The JVLA observations did~not detect the SN in the 22\,GHz and 8.9\,GHz bands
down to a 3-$\sigma$ upper limits of 131\,$\mu$Jy and 65\,$\mu$Jy,
respectively.
The ATCA observations, put a 3-$\sigma$ limit of 66\,$\mu$Jy
in the 18\,GHz band.
Since our previous limits show that there is a significant
amount of CSM interacting with the SN shock in this event,
it is likely that strong synchrotron radiation is generated.
The non-detection of such a radio source implies that $\tau_{{\rm ff}}>1$,
providing, therefore, a lower limit on $\dot{M}$.

\subsection{Precursor fluence}
\label{sec:prec}

Prior to the fast rise detected on 2012 September 24 (Prieto et al. 2012),
the light curve of SN\,2009ip presented a feature which can be
interpreted as $\gtorder 1$\,month-long outburst.
If we assume that this outburst was a mass-loss event
(rather than part of the SN explosion),
and if we assume that the bolometric luminosity of the outburst
is of the same order of magnitude as the kinetic energy released in the outburst,
then by comparing the bolometric luminosity with the kinetic energy
we can get a rough estimate of the mass released in the outburst.
The outburst had a peak absolute $V$-band magnitude of about $-15$,
and a duration of at least 30 days (see Prieto et al. 2012).
Therefore, the total bolometric fluence of the outburst is
$E_{{\rm bol}}\gtorder 8\times10^{47}$\,erg.
The reason for the lower limit is that we do~not know
the outburst light curve bolometric
correction\footnote{The bolometric magnitude correction is always positive.},
and we only have a lower limit on its duration.
Comparing $E_{{\rm bol}}$ with the kinetic energy
and dividing by the duration of the event, $t_{{\rm dur}}$, we get
a lower limit on the mass-loss rate
\begin{eqnarray}
\dot{M} & \gtorder & \frac{2E_{{\rm bol}}}{v^{2}t_{{\rm dur}}} \cr
        & \cong    & 1.6\times10^{-3} v_{{\rm w},2000}^{-2} E_{{\rm bol},8e47} t_{{\rm dur},30}^{-1}\,{\rm M}_{\odot}\,{\rm yr}^{-1},
\label{Ek}
\end{eqnarray}
where $v_{{\rm w},2000}$ is the wind/outburst velocity in units of 2000\,km\,s$^{-1}$, $E_{{\rm bol},8e47}$ is the bolometric energy in units of $8\times10^{47}$\,erg,
and $t_{{\rm dur},30}$ is the outburst duration in units of 30\,days.

\section{Discussion}
\label{sec:Disc}

\subsection{Constraints on Mass loss}
\label{sec:MassLoss}

Along with the observations,
equations~\ref{N}, \ref{eq:KLr}, \ref{Lx}, \ref{tdiff},
\ref{tau_ff} and \ref{Ek}
provide order
of magnitude lower and upper bounds on the mass-loss rate
from the SN progenitor.
It is important to note that the reason these are only
order of magnitude estimates is because some
of the assumptions that go into these
formulae are likely inaccurate.
For example, the assumption that the wind
is infinite, continuous and can be described by a single velocity component,
or the assumption of spherical symmetry.
Nevertheless, these relations provide
order of magnitude, independent, estimators
for the SN progenitor mass-loss rate.
\begin{figure}
\centerline{\includegraphics[width=8.5cm]{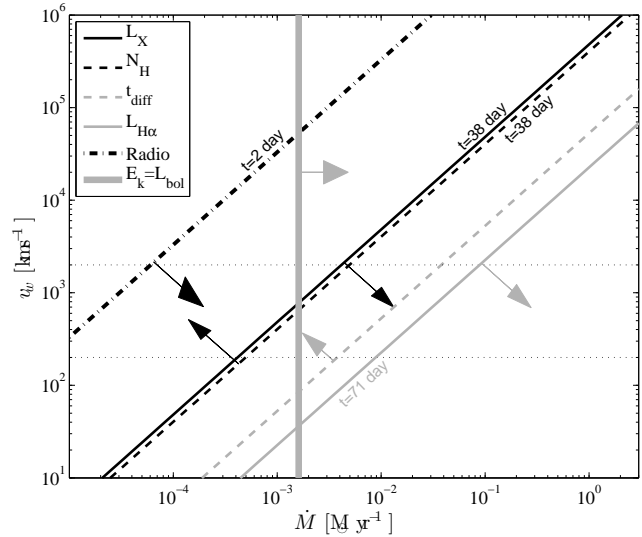}}
\caption{Upper and lower limits on the mass-loss rate
of the SN\,2009ip progenitor as a function of the wind/outburst
velocity.
The assumptions that go into these bounds are discussed in
\S\ref{sec:Form}.
The solid black line shows the limit based on the X-ray luminosity
(Equation~\ref{Lx}).
The solid gray line represents the H$\alpha$ luminosity derived limits
(Equation~\ref{eq:KLr}).
The black dashed line and the gray dashed line
represent the column density (Equation~\ref{N})
and diffusion time scale (Equation~\ref{tdiff}) limits, respectively.
The vertical thick gray solid line is based on the bolometric
fluence (Equation~\ref{Ek}).
Finally, the free-free absorption limit
(Equation~\ref{tau_ff}) is represented by the
black dash-dotted line.
The arrow attached to each line marks the direction of the
region allowed by the line criteria.
The horizontal dotted lines mark the 200 and 2000\,km\,s$^{-1}$
wind/outburst velocity
which we derive from the, presumably pre explosion, IR spectrum
(Figure~\ref{fig:Spec_Ha_PaschenAlpha}).
These lines constitute the approximate range of
plausible wind/outburst velocities.
Since it is likely that the mass-loss was~not a continuous
process with constant mass-loss rate,
these measurements should be regarded as an order of magnitude
estimate. See discussion in \S\ref{sec:Disc}.
\label{fig:MdotVw}}
\end{figure}

Some of the estimators require knowledge regarding
the shock radius $r$.
Following Chevalier (1982), we use the
approximation\footnote{Assuming the power-law index describing the
ejecta velocity distribution is $m=8$. Note that Chevalier (1982) denoted this
variable by $n$, while Balberg \& Loeb (2010) use $m$.}
\begin{eqnarray}
r\sim & \int{v(t)dt}=(5/4)v_{{\rm ej,s}}(t_{{\rm s}}-t_{0})^{1/5}(t-t_{0})^{4/5} \cr
      & \approx 3\times10^{15} \frac{v_{{\rm ej,s}}}{8000\,{\rm km\,s}^{-1}}
        \Big( \frac{t-t_{0}}{30\,{\rm day}} \Big)^{4/5}\,{\rm cm},
\label{r}
\end{eqnarray}
where $t$ is the time, $t_{0}$ is the SN explosion time,
and $v_{{\rm ej,s}}$ is the SN ejecta velocity
($\approx 8000$\,km\,s$^{-1}$) at $t_{{\rm s}}$ ($t_{{\rm s}}-t_{0}=71$\,day),
In Figure~\ref{fig:MdotVw}
we present the limits we derive on $\dot{M}$,
as a function of the wind/outburst velocity.
Specifically,
assuming $N\sim N_{{\rm H}}$,
our X-ray observations of SN\,2009ip provide an upper limit
of $N_{{\rm H}}<2\times10^{22}$\,cm$^{-2}$.
Therefore, Equtaion~\ref{N} constitutes an upper limit on $\dot{M}$
(black dashed line).
Using Equation~\ref{eq:KLr},
the H$\alpha$ line luminosity of the {\it narrow} H$\alpha$ component
we measured on 2012 Dec 4,
$L_{{\rm H}\alpha}\approx1.6\times10^{39}$\,erg\,s$^{-1}$,
and assuming $\beta=1$,
we can set a lower limit on $\dot{M}$ (gray solid line).
As discussed in \S\ref{sec:Form}, Equation~\ref{Lx},
along with our measured X-ray luminosity of
$1.1\times10^{39}$\,erg\,s$^{-1}$
sets a lower limit on $\dot{M}$ which is
shown as the black solid line in Figure~\ref{fig:MdotVw}.
Furthermore, the SN rise time of 2.3\,mag\,day$^{-1}$
(Prieto et al. 2012)
implies $t_{{\rm diff}}\ltorder 0.3$\,day.
Along with Equation~\ref{tdiff} these
provide an upper limit on $\dot{M}$ (gray dashed line).
The estimate based on the bolometric
fluence (Equation~\ref{Ek}) is shown as a vertical gray thick line.
The non detection in radio (Chandra \& Soderberg 2012;
Hancock et al. 2012)
with Equation~\ref{tau_ff} set a lower limit on $\dot{M}$
(black dash-dotted line).
On each line we mark also the $(t-t_{0})$ used to
calculate the line position.
Here we assume that $t_{0}$ is on 2012 Sep 24.

Although there is no single $\dot{M}$ value
which is consistent with all the bounds in Figure~\ref{fig:MdotVw},
the closest values to all the bounds is in the range of
about $10^{-3}$ to $10^{-2}$\,M$_{\odot}$\,yr$^{-1}$.

Kiewe et al. (2012) review the observed properties of
15 type-IIn SNe. They reported mass-loss rates,
prior to explosion,
in the range of $10^{-4}$ to $\sim1$\,M$_{\odot}$\,yr$^{-1}$,
while their wind velocities are in the range of $\sim30$
to 1600\,km\,s$^{-1}$.
The mass-loss rate and wind velocity of
SN\,2009ip is consistent with these values.
Another SN which shows some similarities with
SN\,2009ip is SN\,2010mc (PTF\,10tel; Ofek et al. 2013).
This SN showed an outburst about one month prior
to its explosion. We note that the high state of SN\,2009ip
just prior to its fast rise (Prieto et al. 2012)
can be interpreted as a similar outburst.

Type-IIn SNe are likely a non-homogeneous class of objects
arising from multiple mechanisms. It is not clear what
the best combination of parameters that will help
us to relate a given type-IIn to specific mechanism/progenitor are
(e.g., luminous blue variable eruptions).
However, better mass-loss rates and wind velocity measurements
for larger samples of type-IIn SN progenitors,
as well as additional cases of pre-explosion outbursts
can provide the missing link.

\subsection{Interpretation}
\label{sec:Interp}

With the exception of the mass estimate based on the H$\alpha$
luminosity, the mass-loss estimators in Figure~\ref{fig:MdotVw}
are consistent to an order of magnitude.
Specifically, the mass-loss lower limit based on the H$\alpha$ 
luminosity, and assuming $\beta=1$,
is over an order of magnitude above
the upper limit which is derived from the bound-free absorption
column-density limit.
We note that in the case of another source in which a similar
analysis was applied (SN\,2010mc/PTF\,10tel; Ofek 2012; Ofek et al. 2013b),
the various estimators were consistent.

A possible explanation for the discrepancy here is
that some of our basic assumptions are incorrect.
Among these assumptions, are uniformity of the CSM, spherical symmetry,
$r^{-2}$ density profile,
solar metallicity, and ionized (but not fully ionized)
CSM.
Alternatively,
as we discussed in \S\ref{sec:Ha}, it is possible
that the H$\alpha$ emitting region is further out,
above the shock region (e.g., $\beta\gg1$).
Therefore, it is possible that the H$\alpha$ line luminosity
probes a completely different regions of the CSM,
than the other methods discussed in \S\ref{sec:Form}.

Interestingly, we note that in Figure~\ref{fig:MdotVw},
solution allowed by all the mass-loss rate estimators which depend on
the integral of density along the line of sight
(marked by non-solid lines in Fig.~\ref{fig:MdotVw}),
infer low mass-loss rates,
while estimators which measure the total emission
from an optically thin volume
(marked by solid lines in Fig.~\ref{fig:MdotVw}),
give high values for the mass-loss rate.
This behavior hints that another possible explanation
for the discrepancy between the various lines
in Fig.~\ref{fig:MdotVw} is that
the CSM has an aspherical geometry.

There are two simple geometries that are roughly consistent
with these results.
The first simple explanation is that
the CSM around the SN has a disk geometry,
and we observe the system from above or below
the disk.
In this case there will be a relatively small amount
of intervening material between the observer and
the source, hence small value
of $N_{{\rm H}}$, and short $t_{{\rm diff}}$.
Moreover, in this case the total emission
($L_{{\rm X}}$ and $L_{{\rm H}\alpha}$)
will be larger relative to the expectation
based on the spherical geometry assumption
and on the values of $N_{{\rm H}}$ and $t_{{\rm diff}}$.
The second simple model is that the CSM
has a bipolar hourglass-like structure.
In this case we observe the system
from the equatorial plane.
We note that there are likely other possible geometrical
solutions, which are more complicated.

We conclude that the best explanation
for the discrepancy between the mass-loss estimators
is that the H$\alpha$ emission region
is above the shock region (or effectively $\beta\gg1$),
or alternatively
that the CSM is aspherical.
Unfortunately our order of magnitude analysis
does~not provide a way to distinguish between
the two scenarios.

If the H$\alpha$ emission region is indeed located
further out, relative to the shock,
then an immediate conclusion is that
an order of magnitude estimate to the mass-loss
rate during the SN precursor is in the range of
$\sim10^{-3}$\,M$_{\odot}$\,yr$^{-1}$ to $\sim10^{-2}$\,M$_{\odot}$\,yr$^{-1}$.

In order to convert the mass-loss rate to an estimate of the
total mass in the CSM we need to integrate Equation~\ref{eq:Mr}
out to a specific radius. Here we choose to integrate
the total mass out to radius of $6\times10^{15}$\,cm.
The reason for this choice is that the relatively abrupt disappearance
of the X-ray flux $\sim70$\,days after the explosion
may indicate that the CSM density
is falling (faster than a wind profile) at
a distance of $6\times10^{15}$\,cm (see Equation~\ref{r}).
Using Equation~\ref{eq:Mr} we find that the total CSM mass out
to this radius is
\begin{equation}
M_{{\rm CSM}} \sim 4\times10^{-2} \dot{M}_{0.01} v^{-1}_{{\rm w},500} r_{6E15}\,{\rm M}_{\odot}.
\label{TotalMass}
\end{equation}
Here $r_{6E15}$ is the radius in units of $6\times10^{15}$\,cm.

\subsection{Implications}

Pastorello et al. (2012) and Mauerhan et al. (2012) suggested that
the outbursts of SN\,2009ip are due to pulsational
pair instability.
However, Woosley et al. (2007)
predict that the mass-loss in pair instability mass ejections
would be at least a few solar masses.
Unless the geometry is highly aspherical,
this theoretically predicted mass-loss
is high relative to our estimate of the total mass
in the CSM (i.e., $\sim0.1$\,M$_{\odot}$).
Our mass-loss estimate
is of the same order of magnitude as the one
derived in Ofek et al. (2013b)
in context of the Quataert \& Shiode (2012) mechanism.
The estimators presented in Figure~\ref{fig:MdotVw}
are also in rough agreement
with the shell mass of $\sim0.15$\,M$_{\odot}$
suggested by Soker \& Kashi (2012),
in context of their binary-star merger scenario.

Interestingly, both Levesque et al. (2012)
and Soker \& Kashi (2012) suggested
aspherical models for SN\,2009ip (see also Mauerhan et al. 2012).
Levesque et al. (2012) 
argued for a thin disk geometry,
while Soker \& Kashi (2012)
suggested a bipolar hourglass-like geometry.
However, Soker \& Kashi (2012) suggested
that we are observing the system along the polar
direction.
We note that if the discrepancy in Figure~\ref{fig:MdotVw}
is due to asymmetry in the CSM,
rather than the radius at which the H$\alpha$ line
is generated,
than the Soker \& Kashi (2012) geometry is 
not consistent with our observations.
However,
we do~not claim that our suggested geometries
are the only possible solutions.
Finally, we note that
different mass-loss estimators
have different functional dependencies on $r$.
Therefore, additional observations
(e.g., radio) can constrain the density
profile of the CSM.

\acknowledgments

We thank Orly Gnat, Udi Nakar, Stan Woosley, Nir Sapir and Avishay Gal-Yam
for productive discussions, and an anonymous referee
for useful suggestions.
We gratefully acknowledge the collaboration of the  XMM Project team, and in particular Dr. Norbert Schartel, for the XMM ToO observations. 
E.O.O. is incumbent of
the Arye Dissentshik career development chair and
is grateful to support by
a grant from the Israeli Ministry of Science.
M.M.K. acknowledges generous support from the Hubble Fellowship and
Carnegie-Princeton Fellowship.

\end{document}